\documentclass[notitlepage,english,aps,floats,onecolumn,showpacs,nofootinbib,floatfix]{revtex4-1}

\usepackage{pslatex}
\usepackage[T1]{fontenc}
\usepackage[latin1]{inputenc}
\usepackage{graphicx}
\usepackage{epsfig}
\usepackage{longtable}
\usepackage{float}
\usepackage{calc}
\usepackage{ifthen}
\usepackage{amsmath}
\usepackage{relsize}
\usepackage{hyperref}
\usepackage{amssymb}

\usepackage{color}

{
{
{
\newcommand{\bea}{\begin{eqnarray}}
\newcommand{\eea}{\end{eqnarray}}

\newcommand{\nc}{\newcommand}
\nc{\renc}{\renewcommand}
\nc{\eqs}[2]{\mbox{Eqs.~(\ref{#1},\,\ref{#2})}}
\nc{\eq}[1]{\mbox{Eq.~(\ref{#1})}}
\nc{\figs}[2]{\mbox{Figs.~(\ref{#1},\,\ref{#2})}}
\nc{\fig}[1]{\mbox{Fig~.(\ref{#1})}}
\nc{\be}[1]{\begin{equation} \mbox{$\label{#1}$}}
\nc{\ee}{\vspace{0.1cm}\end{equation}}

\newcommand{\bean}{\begin{eqnarray*}}
\newcommand{\eean}{\end{eqnarray*}}

\def\GeV{{\rm \ GeV}}

\def\lae{\;^{<}_{\sim} \;} \def\gae{\; ^{>}_{\sim} \;}

\begin{document}

\title{Higgs Inflation via a Metastable Standard Model Potential, Generalised Renormalisation Frame Prescriptions and Predictions for Primordial Gravitational Waves}

\author{J.McDonald }
\email{j.mcdonald@lancaster.ac.uk}
\affiliation{Dept. of Physics,  
Lancaster University, Lancaster LA1 4YB, UK}

\begin{abstract}

Higgs Inflation via a metastable Standard Model Higgs Potential is possible if the effective Planck mass in the Jordan frame increases after inflation ends. Here we consider the predictions of this model independently of the dynamics responsible for the Planck mass transition. The classical predictions are the same as for conventional Higgs Inflation. The quantum corrections are dependent upon the conformal frame in which the effective potential is calculated. We generalise beyond the usual Prescription I and II renormalisation frame choices to include intermediate frames characterised by a parameter $\alpha$. We find that the model predicts a well-defined correlation between the values of the scalar spectral index $n_{s}$ and tensor-to-scalar ratio $r$. For values of $n_{s}$ varying between the 2-$\sigma$ Planck observational limits, we find that $r$ varies between 0.002 and 0.005 as $n_{s}$ increases, compared to the classical prediction of 0.003. Therefore significantly larger or smaller values of $r$ are possible, which are correlated with larger or smaller values of $n_{s}$. In addition, the model can be compatible with the larger values of $n_{s}$ predicted by Early Dark Energy solutions to the Hubble tension, with correspondingly larger values of $r$. The model can be tested via the detection of primordial gravitational waves by the next generation of CMB polarisation experiments.   

\end{abstract}
 
\maketitle

\section{Introduction}   

Higgs Inflation \cite{bs} seeks to explain inflation using the only known scalar particle. However, the electroweak vacuum of the Standard Model (SM) may be metastable, due to quantum corrections which cause the Higgs effective potential to become negative at a scale $\phi = \Lambda \sim 10^{10} \GeV$ \cite{frog,sher,unst1,unst2,unst3}
\footnote{In \cite{litimvs} it was found that, using 2023 PDG inputs, there is only a small region of the 4-$\sigma$ ellipse in $m_{t}$ and $\alpha_{s}(M_{z})$ for which stability is possible, and using correlated CMS inputs, which have larger errors but take into account correlations between the measurements, there is only a very small region of the 2-$\sigma$ ellipse for which stability is  possible. In \cite{stein1} it was also found that only a small region of the 2-$\sigma$ ellipse is still compatible with stability. A factor of 2 improvement in the errors will exclude stability to 5-$\sigma$ \cite{litimvs}. However, it should be noted that there is a theoretical uncertainty between the Monte Carlo $t$ quark mass from direct measurements and the pole mass used to calculate the potential,  due to non-perturbative effects that are difficult to quantify \cite{pdg,nason,hoang}. Therefore the likelihood of the metastability of the SM potential is not yet established.}. In this case the Standard Model Higgs potential cannot serve as a basis for Higgs Inflation \footnote{A possible exception to this has been proposed in \cite{nr}, where it is suggested that shifts of the Higgs self-coupling and top quark Yukawa coupling due to finite renormalisation counterterms, whose magnitude depends upon the unknown UV completion, could stabilise the Higgs potential during inflation if they are sufficiently large. Here we are conservatively assuming that such shifts are small.}.      

It is possible to stabilise an otherwise metastable SM Higgs potential, for example by adding scalars with a sufficiently strong coupling to the Higgs boson. However, it is quite possible that such scalars do not exist. In this case Higgs Inflation must use the metastable Higgs potential. The only way that this can work is if the cosmological framework is modified. Specifically, if the effective Planck mass in the Jordan frame increases after the end of inflation, without introducing a significant change in the cosmological energy density in the Einstein frame, then Higgs Inflation via a metastable potential becomes possible. 

In \cite{pt1} we presented a specific model to illustrate that such a Planck mass transition is possible, based on adding a second non-minimally coupled scalar field. In particular, it was shown that the energy density associated with the Planck mass transition can be negligible compared to the energy density from the decay of the inflaton. 
    
In this paper we discuss the general predictions of metastable Higgs Inflation, which are independent of the dynamics of the Planck mass transition. In \cite{pt1} we showed that the classical predictions of the model are exactly the same as for conventional Higgs Inflation. However, quantum corrections can have a strong effect on the predictions of the model. In Higgs Inflation, the form of the quantum corrections to the potential depends upon the conformal frame in which the corrections are calculated. The choice of frame defines the frame in which the renormalisation cut-off is field independent. Since the cut-off represents the scale of the UV completion, the correct renormalisation frame can only be determined once the full UV-completion of the model is known. In the absence of such knowledge, all renormalisation frames are possible\footnote{In \cite{rf1} it is shown that different renormalisation frames correspond to different measures for the path integral of the effective action in a given frame. Knowledge of the UV-completion is necessary to determine the correct measure.}. 

The two commonly considered frame choices are known as  Prescription I, where the effective potential is calculated in the Jordan frame, and Prescription II, where it is calculated in the Einstein frame \cite{p12}. However, in the absence of the UV-complete theory, there is no reason to exclude other frames which are intermediate between the initial Jordan and the final Einstein frame. In such frames there is both a non-minimal coupling of the Higgs field to the Ricci scalar and non-canonical kinetic terms for the SM fields.

Here we will consider an initial transformation to intermediate renormalisation frames ("$\alpha$-frames") defined by a conformal factor $\hat{\Omega} = \Omega^{\alpha}$, where $\Omega$ is the conventional conformal factor of Higgs Inflation. $\alpha = 1$ corresponds to Prescription I and $\alpha = 0$ to Prescription II. We will show that the value of $\alpha$ has a strong effect on the predictions of the model. 

The paper is organised as follows. In Section II we introduce the general framework for our analysis. In Section III we introduce the $\alpha$-frames for calculating the effective potential. In Section IV we calculate the predictions of the model for $n_s$ and $r$. In Section V we show how any renormalisation close to Prescription I can be described by an $\alpha$-frame. In Section VI we present our conclusions. Some details of the calculations are discussed in the appendices.

\section{Metastable Higgs Inflation with a Planck mass transition}

In this section we explain how an increase in the effective Planck mass at the end of inflation can allow inflation with values of the SM Higgs field which are smaller than the instability scale $\Lambda$. We will first consider the classical theory, with the quantum corrections discussed in the following sections.  

The action of the model in the Jordan frame is 
\be{e1}      S = \int d^{4} x \sqrt{-g} \left[ \left(M_{Pl,\,eff}^{2} + \xi_{\phi} \phi^{2} \right) \frac{R}{2}   - \frac{1}{2} \partial_{\mu} \phi \partial^{\mu} \phi  - V(\phi) + {\cal L}_{SM} \right]  ~.\ee
$M_{Pl,\,eff}$ is the Jordan effective Planck mass in the $\phi \rightarrow 0$ limit and ${\cal L}_{SM}$ is the Lagrangian of the SM fields excluding the Higgs kinetic term and potential. 

Metastable Higgs Inflation requires that the following two conditions are satisfied: 

\noindent (i) There is an increase in $M_{Pl,\,eff}$ at or after the end of inflation from $M_{0} \ll M_{Pl}$ to $M_{Pl}$, occurring on a timescale short compared to $H^{-1}$. In this case the transition is effectively instantaneous, with no significant change in the scale factor. 

\noindent (ii) Any energy released by the Planck mass transition is negligible compared to the energy from the decay of the inflaton. 

\noindent If these conditions are satisfied then the model is predictive, independently of the dynamics responsible for the Planck mass transition. 

To analyse inflation and the post-inflation era, we transform the action to the Einstein frame via a conformal transformation to $\tilde{g}_{\mu \nu} = \Omega_{0}^{2} g_{\mu\nu}$, where 
\be{e2} \Omega_{0}^{2} = \left(\frac{M_{Pl,\,eff}^{2}}{M_{Pl}^{2}} + \frac{\xi_{\phi} \phi^{2}}{M_{Pl}^{2}} \right)   ~.\ee 
The Einstein frame action is then 
\be{e3}  S = \int d^{4} x \sqrt{-\tilde{g}} \left[ \frac{M_{Pl}^{2}}{2} \tilde{R} - \frac{3 M_{Pl}^{2}}{4 \Omega_{0}^{4}}{\partial_{\mu} \Omega_{0}^{2} \partial^{\mu} \Omega_{0}^{2} } - \frac{1}{2 \Omega_{0}^{2}} \partial_{\mu}\phi  \partial^{\mu}\phi    - \frac{V(\phi)}{\Omega_{0}^{4}}  + \frac{ {\cal \tilde{L}}_{sm}}{\Omega_{0}^{4}} \right]   ~,\ee
where $\tilde{ {\cal L}}_{SM}$ is the SM Lagrangian in terms of $\tilde{g}_{\mu \nu}$. 
Since $M_{Pl,\,eff}/M_{0}$ is constant before and after the Planck mass transition, the second term can generally be expanded to give 
\be{e4}  S = \int d^{4} x \sqrt{-\tilde{g}} \left[ \frac{M_{Pl}^{2}}{2} \tilde{R} -\frac{1}{2 \Omega_{0}^{2}} \left(1 + \frac{6 \xi_{\phi}^{2} \phi^{2} }{\Omega_{0}^{2} M_{Pl}^{2}} \right) \partial_{\mu} \phi \partial^{\mu} \phi 
 - \frac{V(\phi)}{\Omega_{0}^{4}} + \frac{ {\cal \tilde{L}}_{SM}}{\Omega_{0}^{4}}  \right]   ~.\ee
In the Einstein frame the only effect of the Jordan frame Planck mass transition following Higgs decay and reheating is to change the definition of the canonically normalised fields comprising the radiation. The energy density in radiation is not altered by the Planck transition, since the Planck mass and the metric $\tilde{g}_{\mu \nu}$ in the Einstein frame remain unchanged throughout.

During inflation $M_{pl,\,eff} = M_{0}$. Therefore 
\be{e5} \Omega_{0}^{2} = \left(\frac{M_{0}^{2}}{M_{Pl}^{2}} + \frac{\xi_{\phi} \phi^{2}}{M_{Pl}^{2}} \right)   ~.\ee
To express $S$ during inflation in terms of the conventional conformal factor of Higgs inflation, we define a rescaled Higgs field by 
\be{e6}  \tilde{\phi} = \frac{M_{Pl}}{M_{0}} \phi  ~\ee
 The conformal factor is then 
\be{e7}  \Omega_{0}^{2} = \frac{M_{0}^{2}}{M_{Pl}^{2}}   \left(1  + \frac{\xi_{\phi} \phi^{2}}{M_{0}^{2}} \right) =  \frac{M_{0}^{2}}{M_{Pl}^{2}}   \left(1  + \frac{\xi_{\phi} \tilde{\phi}^{2}}{M_{Pl}^{2}} \right) ~\ee
and action becomes
\be{e8}  S = \int d^{4} x \sqrt{-\tilde{g}} \left[ \frac{M_{Pl}^{2}}{2} \tilde{R} -\frac{1}{2 \Omega^{2}} \left(1 + \frac{6 \xi_{\phi}^{2} \tilde{\phi}^{2} }{\Omega^{2} M_{Pl}^{2}} \right) \partial_{\mu} \tilde{\phi} \partial^{\mu} \tilde{\phi} 
 - \frac{V(\phi)}{\Omega^{4}} \left(\frac{M_{Pl}}{M_{0}}\right)^{4}  + \frac{ \tilde{ {\cal L}}_{SM}}{\Omega^{4}}\left(\frac{M_{Pl}}{M_{0}}\right)^{4}     \right]   ~,\ee
where 
 \be{e9}  \Omega^{2} =  \left(1  + \frac{\xi_{\phi} \phi^{2}}{M_{0}^{2}} \right) \equiv  \left(1  + \frac{\xi_{\phi} \tilde{\phi}^{2}}{M_{Pl}^{2}} \right)  ~. \ee
The classical Higgs potential during inflation is $V(\phi) =\lambda_{\phi} \phi^{4}/4$, so in terms of $\tilde{\phi}$ the Einstein frame action purely for the Higgs boson, $S_{\tilde{\phi}}$, becomes 
\be{e10}  S_{\tilde{\phi}} = \int d^{4} x \sqrt{-\tilde{g}} \left[ \frac{M_{Pl}^{2}}{2} \tilde{R} -\frac{1}{2 \Omega^{2}} \left(1 + \frac{6 \xi_{\phi}^{2} \tilde{\phi}^{2} }{\Omega^{2} M_{Pl}^{2}} \right) \partial_{\mu} \tilde{\phi} \partial^{\mu} \tilde{\phi} 
 - \frac{V(\tilde{\phi})}{\Omega^{4}}    \right]   ~.\ee
This action is identical in form to that of conventional Higgs inflation but it is expressed in terms of $\tilde{\phi}$, which is related to the SM Higgs field by $\phi = (M_{0}/M_{Pl}) \tilde{\phi}$. So a large value of $\tilde{\phi}$ can correspond to a small value of the SM Higgs field $\phi$. In particular, it is possible to have $\phi < \Lambda$ if $M_{0}/M_{Pl}$ is small enough. Thus inflation can be achieved using a metastable SM Higgs potential. 
 
It also follows that the predictions of the model when expressed in terms of the number of e-foldings of inflation $N$ are the same as in conventional Higgs Inflation. Therefore, provided that the value of $N$ at the pivot scale is the same in conventional Higgs inflation, the predictions of the model will be the same as conventional Higgs Inflation. This will generally be satisfied, since by condition (ii) the energy density prior to and after the Planck mass transition is the same. Therefore the temperature of the Universe will be unchanged by the Planck mass transition; all that changes in the Einstein frame is the definition of the canonically normalised fields which form the radiation background\footnote{In particular, the massless gauge bosons are conformally invariant and therefore do not change  when $\Omega$ changes due to the Planck mass transition.}. Thus the model will be indistinguishable from conventional Higgs inflation in both its inflation and post-inflation evolution, with the same value of $N$ at the pivot scale and the same predictions for inflation observables. 
The only difference is in the value of the canonically normalised Higgs field during inflation relative to that of the SM Higgs field, which allows inflation with a small value of $\phi$.

\section{Generalised Renormalisation Frames and Quantum Corrections}

In the case of the non-minimally coupled Higgs scalar, the calculation of the quantum-corrected effective potential depends upon the conformal frame in which the theory is renormalised. The correct frame requires knowledge of the UV-completion of the theory, therefore from the point of view of the low-energy effective theory all frames are possible.  There are two renormalisation frames considered in the literature, known as Prescription I and Prescription II \cite{p12}. Prescription I first transforms the classical theory to the Einstein frame and then computes the effective potential. Prescription II first calculates the effective potential in the Jordan frame and then transforms the complete effective potential to the Einstein frame. These are not physically equivalent. In particular, in Prescription I the $\phi$-dependence of the logarithms of the 1-loop Coleman-Weinberg (CW) correction becomes suppressed at $\phi \gae M_{Pl}/\sqrt{\xi_{\phi}}$. As a result, the effect of quantum corrections is milder in Prescription I than in Prescription II.

Prescriptions I and II are special cases of an infinite number of possible renormalisation frames, any one of which could be the correct frame for calculating the effective potential. The Jordan frame used in Prescription II is the special case in which there is a non-minimal coupling of the Higgs to gravity but the SM fields have canonical kinetic terms. At the other extreme, the Einstein frame used in Prescription I is the limit where the Higgs is minimally coupled to gravity but the SM fields have non-canonical kinetic terms. However, there are an infinite number of alternatives, in which there is both a non-minimal coupling of the Higgs to gravity and non-canonical kinetic terms for the SM fields. In the following we will consider the effects of a generalised renormalisation frame on the predictions of the model.  

\subsection{$\alpha$-frames and the Calculation of the Effective Potential}

We consider a family of renormalisation frames defined by a conformal transformation from the Jordan frame with conformal factor $\hat{\Omega}$, where
\be{qc1} \hat{\Omega}^{2} = \Omega^{2 \alpha} ~,\ee
where $\Omega$ is given by \eq{e8}. 
The transformation from the Jordan frame to the final Einstein frame therefore occurs in two stages:
\be{qc2}  g_{\mu\,\nu} \rightarrow \hat{g}_{\mu\,\nu} = \Omega^{2 \alpha}g_{\mu\,\nu}    ~,\ee
followed by    
\be{qc3}  \hat{g}_{\mu\,\nu} \rightarrow \overline{g}_{\mu\,\nu} = \Omega^{2 - 2 \alpha}\hat{g}_{\mu\,\nu} \equiv \Omega^{2} g_{\mu\,\nu}   ~,\ee
with the quantum corrections being calculated in the intermediate frame.  
This gives us a simple one parameter family of frames which contains the Prescription I and Prescription II frames as special cases ($\alpha = 1$ and $\alpha = 0$, respectively). We will refer to these intermediate renormalisation frames as $\alpha$-frames.

Since $\Omega \rightarrow 1$ when $\phi < M_{0}/\sqrt{\xi_{\phi}}$, we can safely run the SM renormalisation group (RG) equations up to a scale $\mu_{c}$ close to $M_{0}/\sqrt{\xi_{\phi}}$, since up to this scale the SM fields have canonical kinetic terms and the metric is the Minkowski metric.  We can then use the 1-loop CW potential in terms of the masses squared of the canonically normalised SM fields in the intermediate frame to compute the effective potential at values of $\phi$ where $\Omega > 1$, as long as the logarithms in the potential are not very large. We finally transform the complete effective potential to the Einstein frame in which inflation is analysed.

  As we are calculating the effective potential during inflation, we can set $M_{Pl,\,eff} = M_{0}$. We start from the Jordan frame action 
\be{qc4}      S = \int d^{4} x \sqrt{-g} \left[ \left(M_{0}^{2} + \xi_{\phi} \phi^{2} \right) \frac{R}{2}   - \frac{1}{2} \partial_{\mu} \phi \partial^{\mu} \phi  - V(\phi) + {\cal L}_{SM} \right]  ~.\ee
We then transform to the $\alpha$-frame via the conformal factor 
\be{qc5}  \hat{\Omega}^{2} = \Omega^{2 \alpha} = \left( 1 + \frac{\xi_{\phi} \phi^{2}}{M_{0}^{2}} \right)^{\alpha}   ~.\ee
The action in the $\alpha$-frame is then 
\be{qc6}  S = \int d^{4} x \sqrt{-\hat{g}} \left[ \frac{M_{0}^{2}}{2} \Omega^{2 -  2 \alpha}\hat{R} -\frac{1}{2 \hat{\Omega}^{2 \alpha}} \left(1 + \frac{6 \xi_{\phi}^{2} \phi^{2} }{\hat{\Omega}^{2 \alpha} M_{0}^{2}} \right) \partial_{\mu} \phi \partial^{\mu} \phi 
 - \frac{V(\phi)}{\hat{\Omega}^{4 \alpha}} + \frac{{\cal \hat{L}}_{SM}}{\hat{\Omega}^{4 \alpha}}   \right]   ~,\ee
where $\hat{g}_{\mu \nu} = \Omega^{2 \alpha} g_{\mu \nu}$. 
To obtain the canonically normalised SM fields, with the exception of the Higgs boson, the scalars and fermions are rescaled according to 
$\hat{\phi} = \phi/\Omega^{\alpha}$ and $\hat{\psi} = \psi/\Omega^{3 \alpha/2}$. The masses of the canonically normalised SM fields, $\hat{M}$, are then related to the Jordan frame mass terms, $M_{J}(\phi)$, by
\be{qc7}  \hat{M}(\phi) = \frac{ M_{J}(\phi)}{\Omega^{\alpha}}   ~.\ee 
The ${\rm \overline{MS}}$ scheme 1-loop CW potential in the $\alpha$-frame at $\mu = \mu_{c}$ is given by 
\be{qc7a} \Delta \hat{V}_{1-loop} = \mathlarger{\sum}_{i} \;\frac{C_{i} \hat{M}_{i}^{4}}{64 \pi^{2}} \left[\ln\left(\frac{\hat{M}_{i}^{2}}{\mu_{c}^{2}}\right) - K_{i}\right]   ~,\ee
where $(C_{i}, K_{i})  = (3, 3/2)$ for the Goldstone bosons, (6, 5/6) for the $W$ bosons, (3, 5/6) for the $Z$ boson, and (-12, 3/2) for the t-quark. In these we have summed over the 3 Goldstone bosons, 2 $W$ bosons and all t-quark colours. We do not include the physical SM Higgs boson as its contribution is suppressed by the non-minimal 
coupling \cite{wilczek,corr1,corr2,barv}. The terms in the 1-loop CW potential in the $\alpha$-frame therefore have the form   
\be{qc8} \Delta \hat{V}_{1-loop} \sim \frac{C \,M_{J}^{4}}{16 \pi^{2} \Omega^{4 \alpha}} \left( \ln \left(\frac{M_{J}^{2}(\phi)}{\mu_{c}^{2} \Omega^{2 \alpha} }\right) - K \right)  ~.\ee 
We then transform to an Einstein frame in which the Einstein-Hilbert term is expressed in terms of $M_{0}$, via a conformal factor $\Omega^{2 - 2 \alpha}$. This gives for the action of the Higgs scalar 
 \be{qc9}  S_{\phi} = \int d^{4} x \sqrt{-\overline{g}} \left[ \frac{M_{0}^{2}}{2} \overline{R} -\frac{1}{2 \Omega^{2}} \left(1 + \frac{6 \xi_{\phi}^{2} \phi^{2} }{\Omega^{2} M_{0}^{2}} \right) \partial_{\mu} \phi \partial^{\mu} \phi 
 - \overline{V}_{E}(\phi) \right] ~,\ee 
where $\overline{g}_{\mu \nu} = \Omega^{2 - 2 \alpha} \hat{g}_{\mu \nu} = \Omega^{2} g_{\mu\nu}$ and 
\be{qc10} \overline{V}_{E}(\phi) = \frac{V(\phi)}{\Omega^{4}}  + \frac{C\; M_{J}^{4}}{16 \pi^{2} \Omega^{4}} \left( \ln \left(\frac{M_{J}^{2}(\phi)}{\mu_{c}^{2} \Omega^{2 \alpha} }\right) - K \right)  ~.\ee
Finally, to analyse inflation, we rescale the metric by a constant factor to $\tilde{g}_{\mu \nu} = (M_{0}^{2}/M_{Pl}^{2})  \overline{g}_{\mu \nu}$ and rescale the Higgs field to 
$\tilde{\phi} = (M_{Pl}/M_{0}) \phi$, which gives the action
for the Higgs scalar $\tilde{\phi}$ in terms of a conventional Planck mass Einstein-Hilbert term
 \be{qc11} S_{\tilde{\phi}} = \int d^{4} x \sqrt{-\tilde{g}} \left[ \frac{M_{Pl}^{2}}{2} \tilde{R} -\frac{1}{2 \Omega^{2}} \left(1 + \frac{6 \xi_{\phi}^{2} \tilde{\phi}^{2} }{\Omega^{2} M_{Pl}^{2}} \right) \partial_{\mu} \tilde{\phi} \partial^{\mu} \tilde{\phi} 
 - V_{E}(\tilde{\phi}) \right] ~,\ee
where
\be{qc12} V_{E}(\tilde{\phi}) =  \left(\frac{M_{0}}{M_{Pl}}\right)^{4} \overline{V}_{E}(\phi) ~.\ee
Therefore
\be{qc13} V_{E}(\tilde{\phi}) = \frac{V(\tilde{\phi})}{\Omega^{4}}  + \frac{C M_{J}^{4}(\tilde{\phi})}{16 \pi^{2} \Omega^{4}} \left( \ln \left(\frac{M_{J}^{2}(\tilde{\phi})}{\tilde{\mu}_{c}^{2} \Omega^{2 \alpha} }\right) - K \right)  ~,\ee
where $M_{J}(\tilde{\phi})$ has $\tilde{\phi}$ in place of $\phi$ and $\tilde{\mu}_{c} = (M_{Pl}/M_{0}) \mu_{c} $. All couplings are calculated at the renormalisation scale $\mu_{c}$.  In writing \eq{qc13} we have assumed that $M_{J}(\phi) \propto \phi$, which is true for all SM particles at large $\phi$.

\section{Predictions for the scalar spectral index and primordial gravitational waves}  

\subsection{Method}

We next discuss the predictions of the metastable Higgs Inflation model for $n_{s}$ and $r$ as a function of $\alpha$. To do this we first run the 2-loop SM RG equations, extended to include the non-minimal coupling $\xi_{\phi}$, from $\mu_{0} = m_{t}$ to $\mu_{c} = 0.1 M_{0}/\sqrt{\xi_{\phi}}$. Since we are calculating the effective potential, the renormalisation scale is equal to the Higgs field $\phi$. We choose to run the RG equations up to this value of $\mu_{c}$ since for $\phi \leq \mu_{c}$ we have $\Omega \approx 1$ and so the RG equations for the SM couplings are the same in all frames, whilst the value of $\phi$ during inflation, $\phi \sim \sqrt{N} M_{0}/\xi_{\phi}$, is small enough compared to $\mu_{c}$ to not cause large logarithms in the CW potential. The RG equations for the non-minimally coupled SM are same as the conventional SM RG equations, up to a modification due to the suppression of the contribution of the physical Higgs scalar at $\mu \equiv \phi > M_{0}/\xi_{\phi}$ due to kinetic term mixing of the physical Higgs scalar with the graviton. This causes a suppression of the physical Higgs propagator by a factor $s(\phi)$, where \cite{wilczek,corr1,corr2,barv}
\be{qc14} s(\phi) = \frac{1 + \frac{\xi_{\phi} \phi^{2}}{M_{0}^{2}} }{ 1 + (1 + 6 \xi_{\phi})\frac{\xi_{\phi} \phi^{2}}{M_{0}^{2}}  } ~. \ee
Therefore we can apply the SM RG equations, with the physical Higgs contribution to the RG equations suppressed by the factor $s(\phi)$. We use the 1-loop RG equations for the non-minimally coupled SM including the $s(\phi)$ factors from  \cite{corr1,corr2} (given in Appendix B) and the 2-loop corrections from \cite{wilczek}. The RG evolution is dominated by the 1-loop terms.   

Once we have determined the SM couplings and $\xi_{\phi}$ at the renormalisation scale $\mu_{c}$, we compute the 1-loop CW potential in the $\alpha$-frame, using $\overline{{\rm MS}}$ scheme SM inputs for the RG equations at $\mu = m_{t}$. The 1-loop correction to the potential in the $\alpha$-frame is 
$$ \Delta \hat{V}_{1-loop} = \frac{1}{64 \pi^{2}} 
\left[ 
 3 \left(\frac{\lambda_{\phi} \phi^{2}}{ \Omega^{2 \alpha}} \right)^{2} \left( \ln \left(\frac{\lambda_{\phi} \phi^{2} }{\mu^{2} \Omega^{2 \alpha} } \right) -\frac{3}{2} \right) 
- 12 \left(\frac{y_{t}^{2} \phi^{2}}{2\Omega^{2 \alpha} } \right)^{2} \left( \ln \left(\frac{y_{t}^{2} \phi^{2}  }{2 \mu^{2} \Omega^{2 \alpha} } \right) -\frac{3}{2} \right) 
\right. $$
\be{qc15} \left. 
+ 6 
\left(\frac{g_{2}^{2} \phi^{2}}{4\Omega^{2 \alpha} } \right)^{2} \left( \ln \left(\frac{g_{2}^{2} \phi^{2}}{4 \mu^{2} \Omega^{2 \alpha} } \right) -\frac{5}{6} \right) 
+ 3 
\left(\frac{(g_{2}^{2}+g_{1}^{2}) \phi^{2}}{4\Omega^{2 \alpha} } \right)^{2} \left( \ln \left(\frac{(g_{2}^{2}+ g_{1}^{2}) \phi^{2}}{4 \mu^{2} \Omega^{2 \alpha} } \right) -\frac{5}{6} \right) \right]   ~,\ee
where the contributions are from the Goldstone bosons, the t-quark and the W and Z bosons, respectively, with the physical Higgs contribution being suppressed by the non-minimal coupling. 
$g_{i}$ are the gauge couplings and $y_{t}$ is the top quark Yukawa coupling. 
After the change of frame to the final Einstein frame, the Einstein potential in terms of $\phi$ becomes
\be{qc16} \overline{V}_{E}(\phi) =  \frac{V(\phi)}{\Omega^{4}}  + \frac{\Delta \hat{V}_{1-loop} }{\Omega^{4 - 4 \alpha} }    ~,\ee
where all couplings are calculated at the renormalisation scale $\mu_{c} = 0.1 M_{0}/\sqrt{\xi_{\phi}}$. The potential $V_{E}(\tilde{\phi})$ is finally obtained by rescaling $\overline{V}_{E}(\phi)$ as in \eq{qc12}.

For the $\overline{{\rm MS} }$ inputs at the renormalisation scale $\mu = m_{t} = 173.1$ GeV, we use the values listed in \cite{martin}: $ v = 246.60109$ GeV, $\lambda_{h} = 0.12604$, $g_{3} = 1.16362$, $g_{2} = 0.64766$, $g_{1} = 0.35854$, $y_{t} = 0.93480$. $\xi_{\phi}(m_{t})$ is a free parameter of the model, which we choose by adjusting the amplitude of the power spectrum at the pivot scale to be equal to its observed value,  $A_{s} =  2.1 \times 10^{-9} $. In our analysis we assume that the pivot scale corresponds to $N = 57$ e-foldings, which is a good estimate for instant reheating after slow-roll inflation (Appendix A).

The analysis of inflation is conducted in the Einstein frame with canonically normalised inflaton $\sigma$, where 
\be{qc17}  \left(\frac{d \sigma}{d \tilde{\phi}}\right)^{2} =  \frac{1}{ \Omega^{2}} \left(1 + \frac{6 \xi_{\phi}^{2} \tilde{\phi}^{2} }{\Omega^{2} M_{Pl}^{2}} \right) ~.\ee
The number of e-foldings of slow-roll inflation at a given $\sigma$ is obtained numerically from
\be{qc18}  N(\sigma) = -\frac{1}{M_{Pl}^{2}} \int_{\sigma}^{\sigma_{end}} \frac{V_{E}}{V_{E}'} d \sigma   ~,\ee
where $\sigma_{end}$ is defined to be the value at which either $|\eta|$ or $\epsilon$ first become greater than 1 as $\sigma$ decreases, with the slow-roll parameters defined with respect to $\sigma$. The scalar spectral index and the tensor-to-scalar ratio are  computed in the usual way, $n_{s} = 1 + 2 \eta - 6 \epsilon$ and $r = 16 \epsilon$, with $\eta = M_{Pl}^{2}(V^{''}_{E}(\sigma)/V_{E}(\sigma))$ and $\epsilon = (M_{Pl}^{2}/2)(V^{'}_{E}(\sigma)/V_{E}(\sigma))^{2}$, where primes denote derivatives with respect to $\sigma$, and the amplitude of the power spectrum is $ A_{s} = V_{E}^{3}/(12 \pi^{2} M_{Pl}^{6} V_{E}^{'\,2})$.  We numerically compute the values of $n_{s}$ and $r$ by using the potential \eq{qc13} in terms of $\tilde{\phi}$ and relating the potential and its derivatives with respect to $\tilde{\phi}$ to the corresponding quantities in terms of $\sigma$ via \eq{qc17}, with the values of $n_{s}$ and $r$ being calculated at $\tilde{\phi}(N = 57)$.

\subsection{Results} 

For reference, we show in Figure 1 the conventional SM Higgs potential using our SM inputs.  The SM potential has an instability at $\phi = \Lambda = 4.2 \times 10^{10} \GeV$.

\begin{figure}[h]
\begin{center}
\hspace*{-0.5cm}\includegraphics[trim = -3cm 0cm 0cm 0cm, clip = true, width=0.55\textwidth, angle = -90]{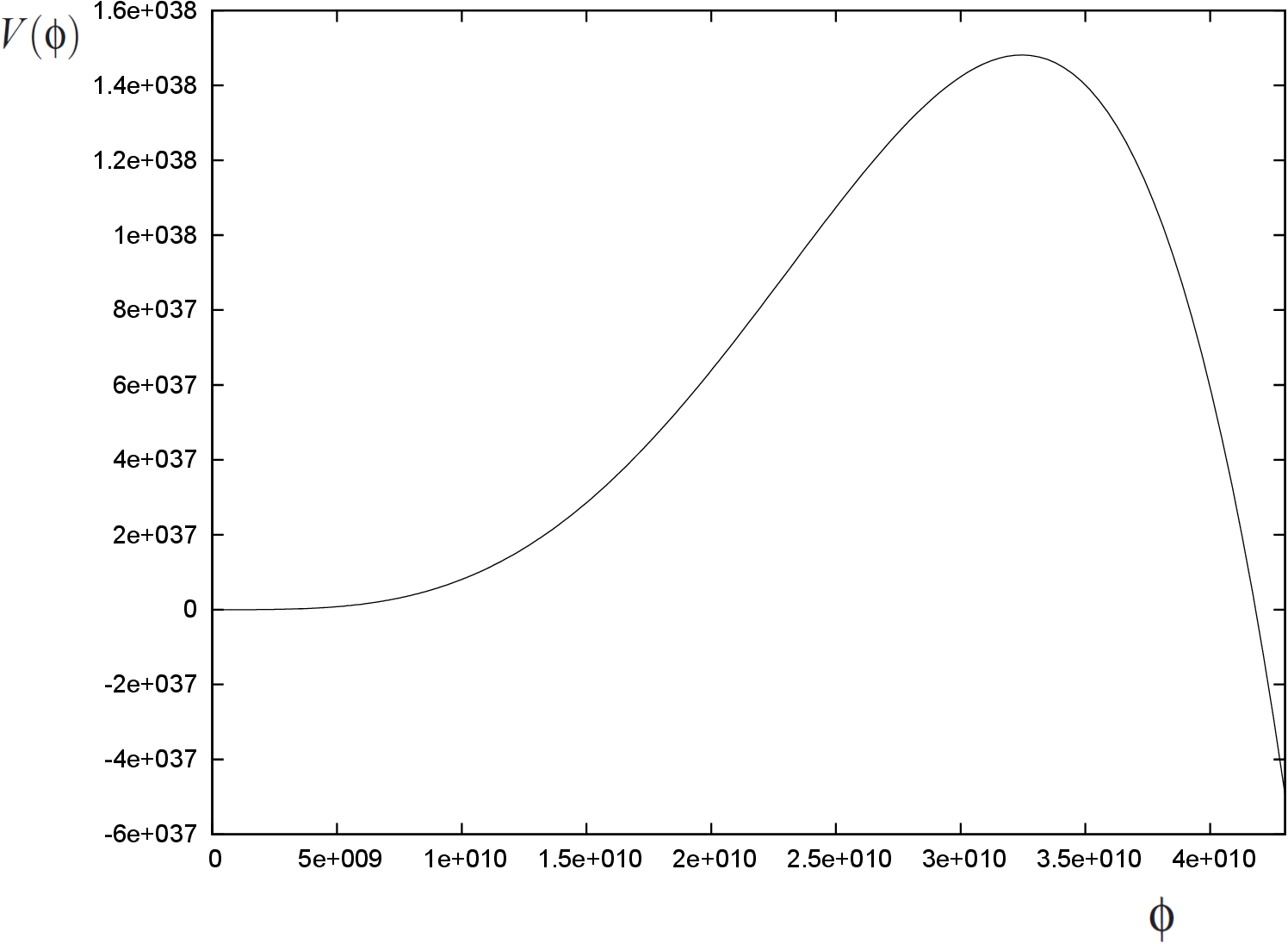}
\caption{The unmodified SM Higgs potential.} 
\label{fig1}
\end{center}
\end{figure}

\begin{figure}[h]
\begin{center}
\hspace*{-0.5cm}\includegraphics[trim = -3cm 0cm 0cm 0cm, clip = true, width=0.55\textwidth, angle = -90]{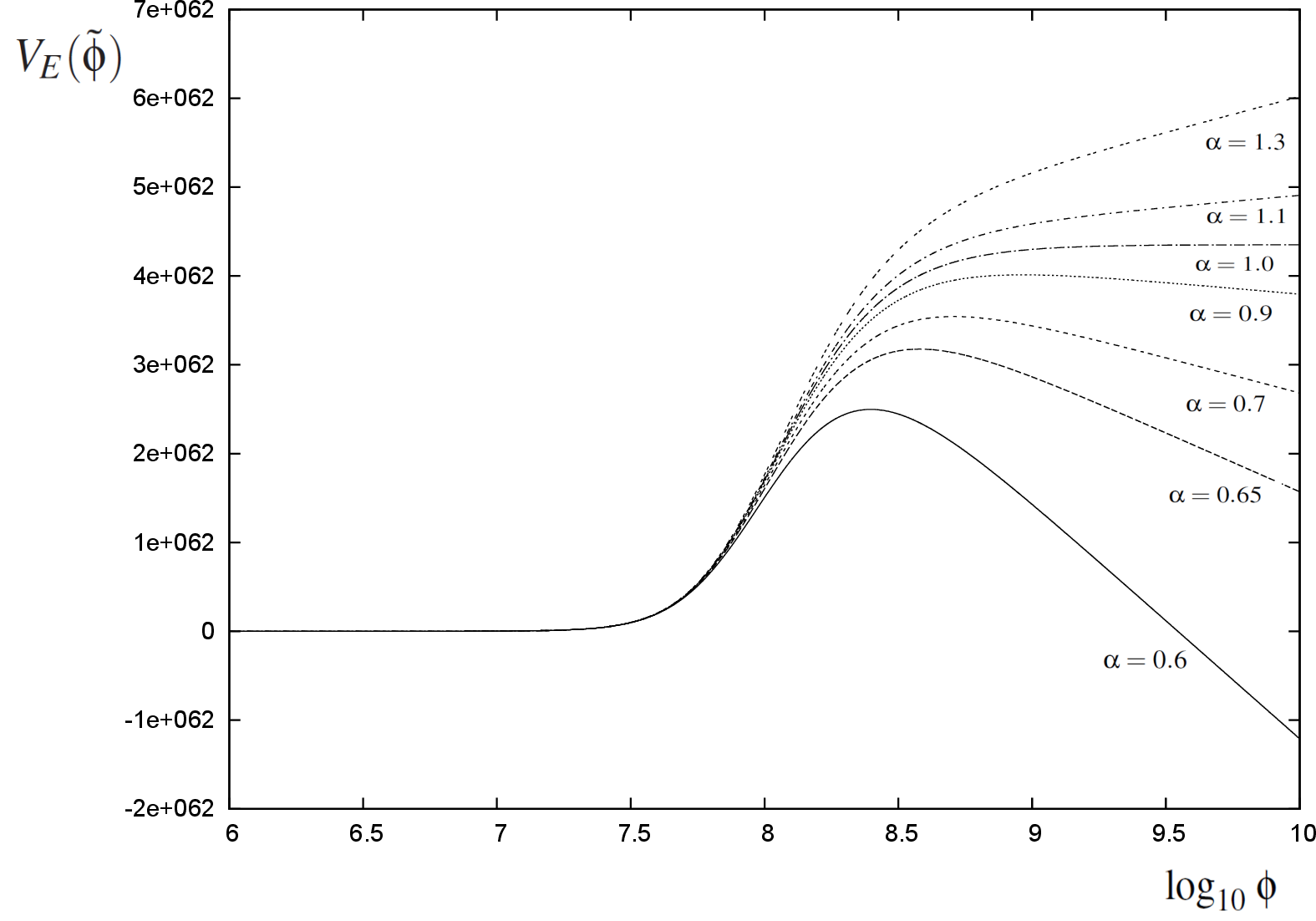}
\caption{Einstein frame potential $V_{E}(\tilde{\phi})$ for different $\alpha$ in the range 0.6 to 1.3 for $M_{0} = 10^{10} \GeV$ and fixed $\xi_{\phi}(m_{t})$. In this we have shown the potential as a function of the SM Higgs $\phi$, with the corresponding value of $\tilde{\phi}$ given by $\tilde{\phi} = (M_{Pl}/M_{0}) \phi$.} 
\label{fig2}
\end{center}
\end{figure}

\begin{figure}[h]
\begin{center}
\hspace*{-0.5cm}\includegraphics[trim = -3cm 0cm 0cm 0cm, clip = true, width=0.6\textwidth, angle = 90]{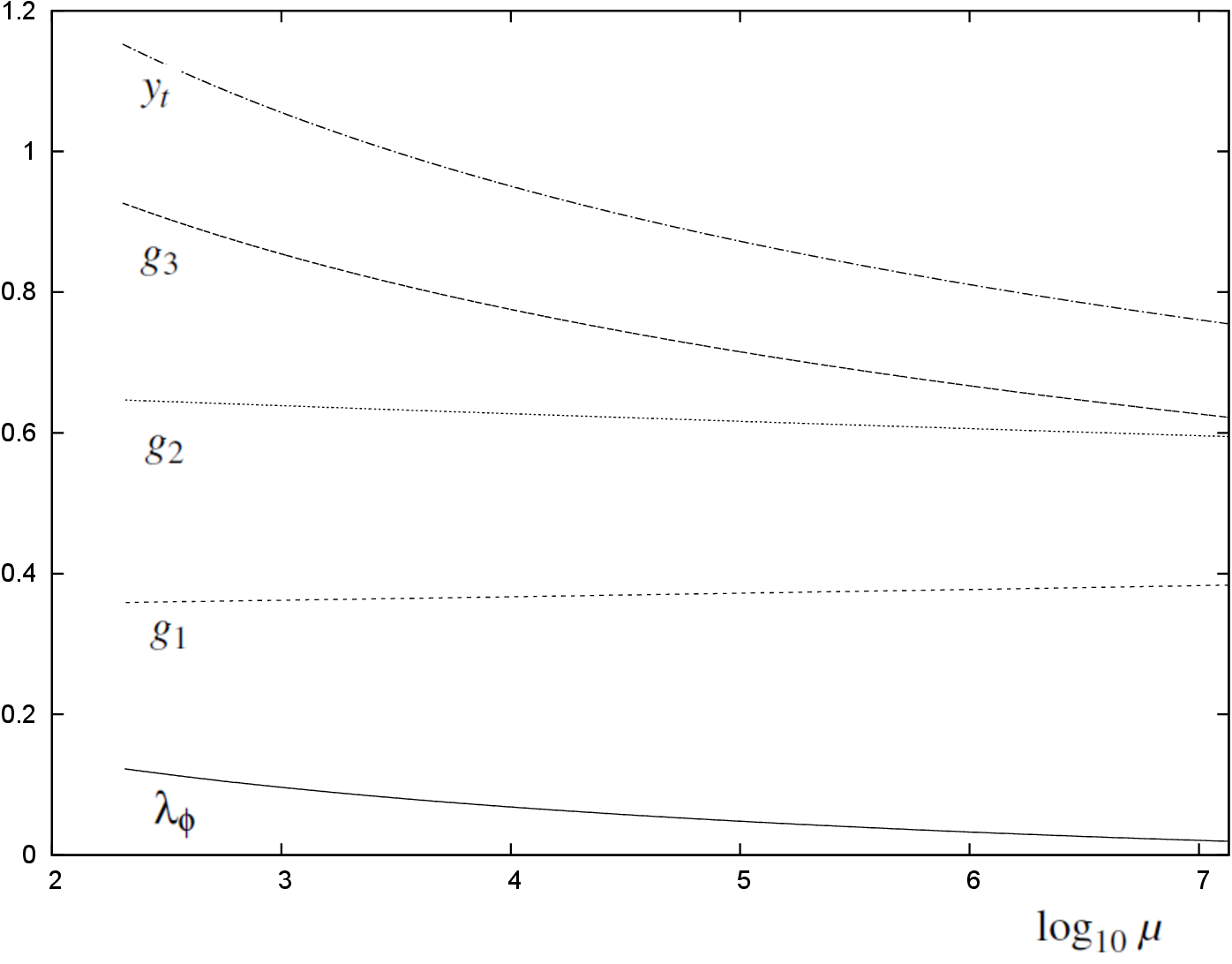}
\caption{SM couplngs for $\mu$ running from $m_{t}$ to $\mu_{c}$, for metastable Higgs Inflation with $M_{0} = 10^{10} \GeV$ and $\alpha = 1$.} 
\label{fig3}
\end{center}
\end{figure}

In Figure 2 we show how the Einstein frame potential $V_{E}(\tilde{\phi})$  varies as a function of $\alpha$ for the case with $M_{0} = 10^{10} \GeV$. In this figure we have set the value of $\xi_{\phi}(m_{t})$ to be equal in all models, in order to compare their behaviour. For realistic potentials the value of $\xi_{\phi}(m_{t})$ should be adjusted for each $\alpha$ to give the correct value of $A_{s}$. We have shown the potential as a function of the SM Higgs $\phi$ rather than of $\tilde{\phi}$, with the corresponding value of $\tilde{\phi}$ given by \eq{e6}.    
For $\alpha < 1$ the potential develops a maximum and then decreases as $\phi$ increases. Nevertheless, inflation with $\alpha < 1$ is still possible if $\phi(N = 57)$ is to the left of the maximum of the potential. In contrast, for $\alpha \geq 1$ the potential is monotonically increasing with $\phi$. Therefore there is no need to carefully choose the initial value of $\phi$ in order to achieve inflation in this case.  

In Figure 3 we show the RG evolution of the SM couplings from $\mu = m_{t}$ to $\mu = \mu_{c}$ for the non-minimally coupled model with $M_{0} = 10^{10} \GeV$.

In Table 1 we show our results for $n_{s}$ and $r$ as a function of $\alpha$ for $M_{0} = 10^{10} \GeV$. We also show the values of $\phi(N = 57)$, $\xi_{\phi}(m_{t})$, $\xi_{\phi}(\mu_{c})$ and $\lambda_{\phi}(\mu_{c})$. In Figure 4 we show $n_{s}$ as a function of $\alpha$ for $M_{0} = 10^{10}\GeV$. We also show the Planck 2-$\sigma$ bounds on $n_{s}$. In Figure 5 we show $r$ as a function of $\alpha$, with the bounds on $\alpha$ corresponding to the 2-$\sigma$ Planck bounds on $n_{s}$ indicated. In Figure 6 we show $r$ as a function of $n_{s}$ for $M_{0} = 10^{10} \GeV$. The model predicts a specific correlation of the values of $n_{s}$ and $r$, with $r$ rapidly increasing as $n_{s}$ increases.

\begin{figure}[h]
\begin{center}
\hspace*{-0.5cm}\includegraphics[trim = -3cm 0cm 0cm 0cm, clip = true, width=0.55\textwidth, angle = -90]{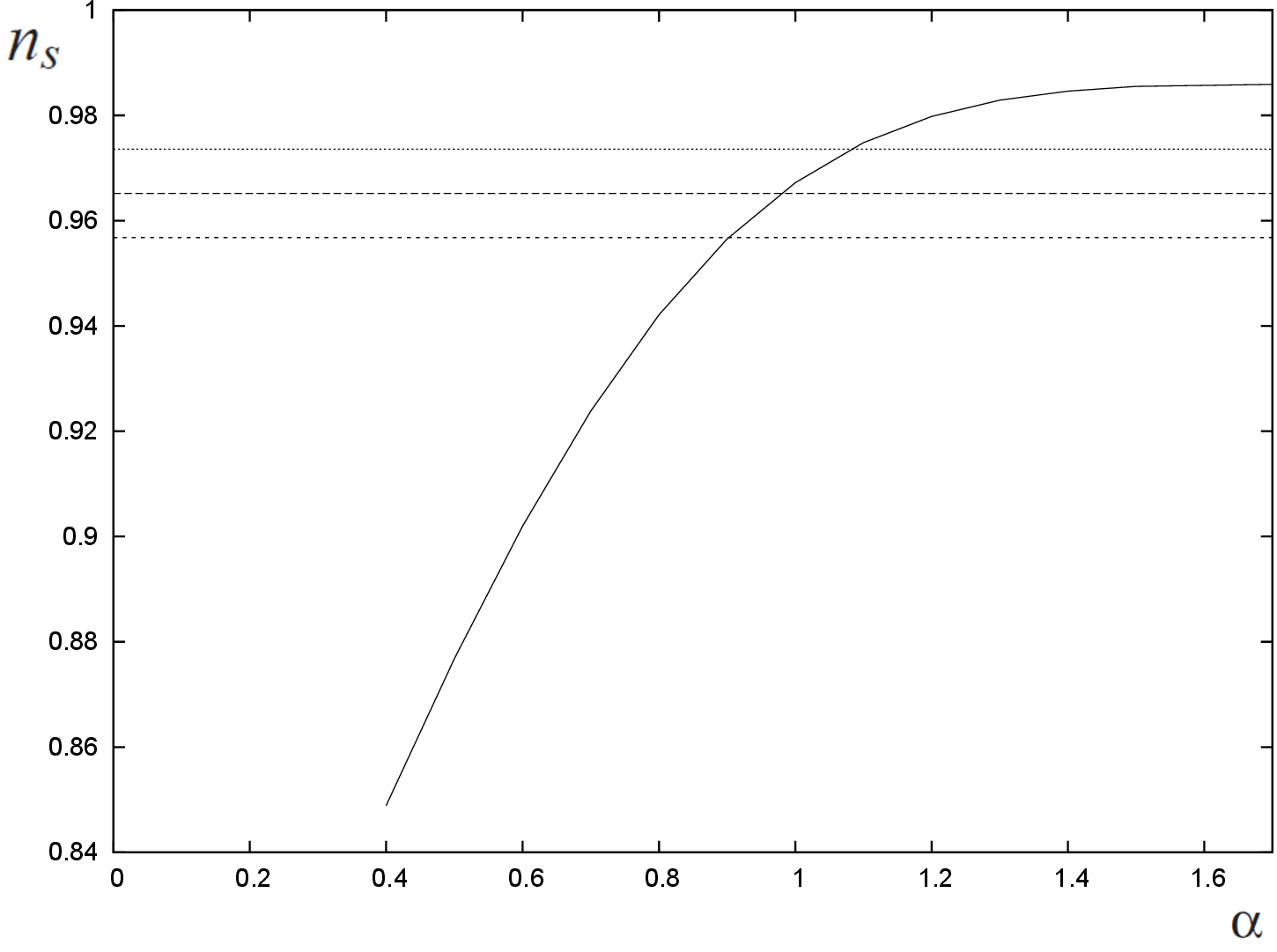}
\caption{$n_{s}$ as a function of $\alpha$ for $M_{0} = 10^{10} \GeV$. The Planck best-fit value  and 2-$\sigma$ upper and lower bounds on $n_{s}$ are indicated.} 
\label{fig4}
\end{center}
\end{figure}

\begin{figure}[h]
\begin{center}
\hspace*{-0.5cm}\includegraphics[trim = -3cm 0cm 0cm 0cm, clip = true, width=0.55\textwidth, angle = -90]{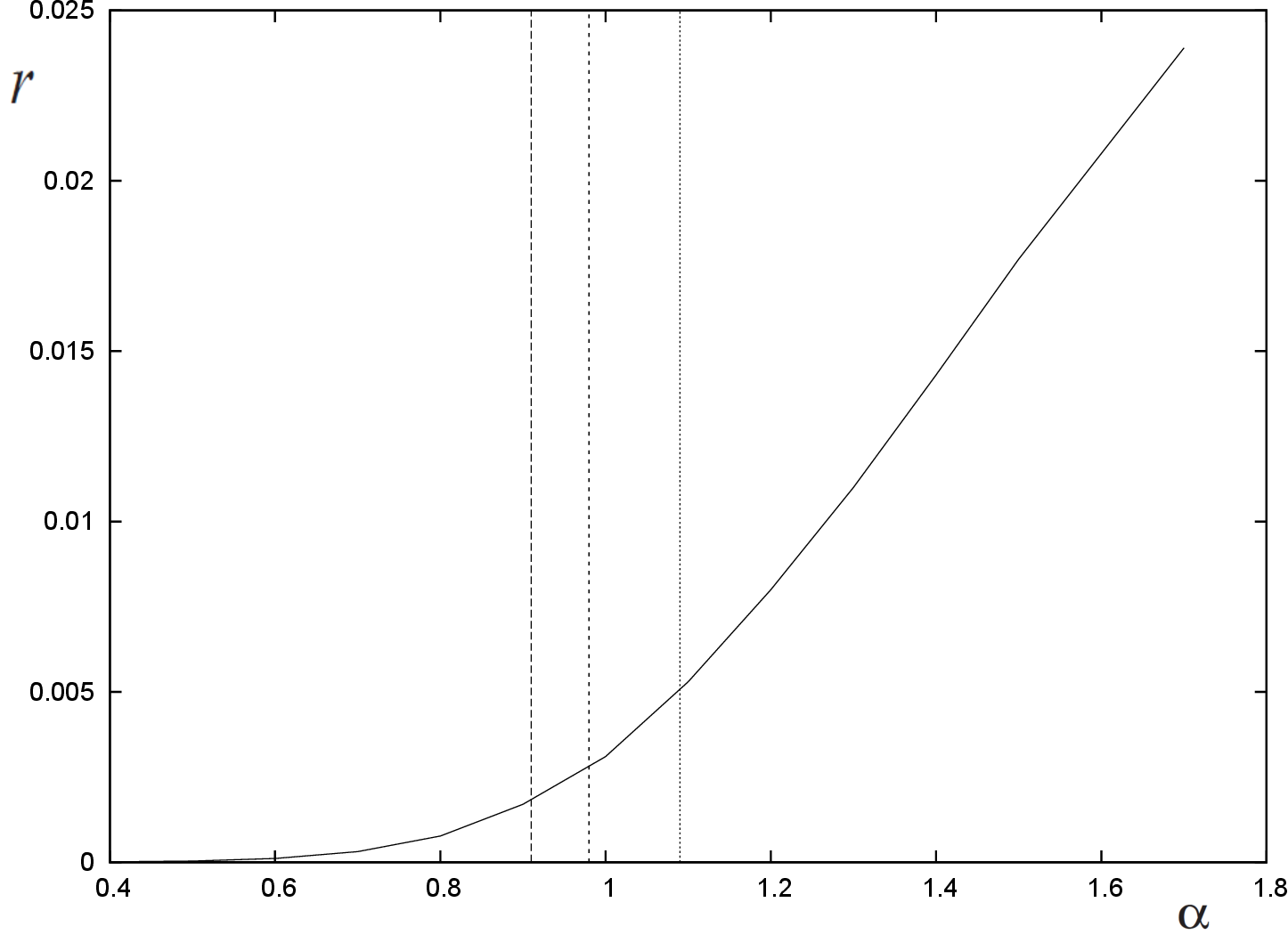}
\caption{$r$ as a function of $\alpha$ for $M_{0} = 10^{10} \GeV$. The Planck best-fit value and 2-$\sigma$ upper and lower bounds from $n_{s}$ are indicated.} 
\label{fig5}
\end{center}
\end{figure} 

\begin{figure}[h]
\begin{center}
\hspace*{-0.5cm}\includegraphics[trim = -3cm 0cm 0cm 0cm, clip = true, width=0.55\textwidth, angle = -90]{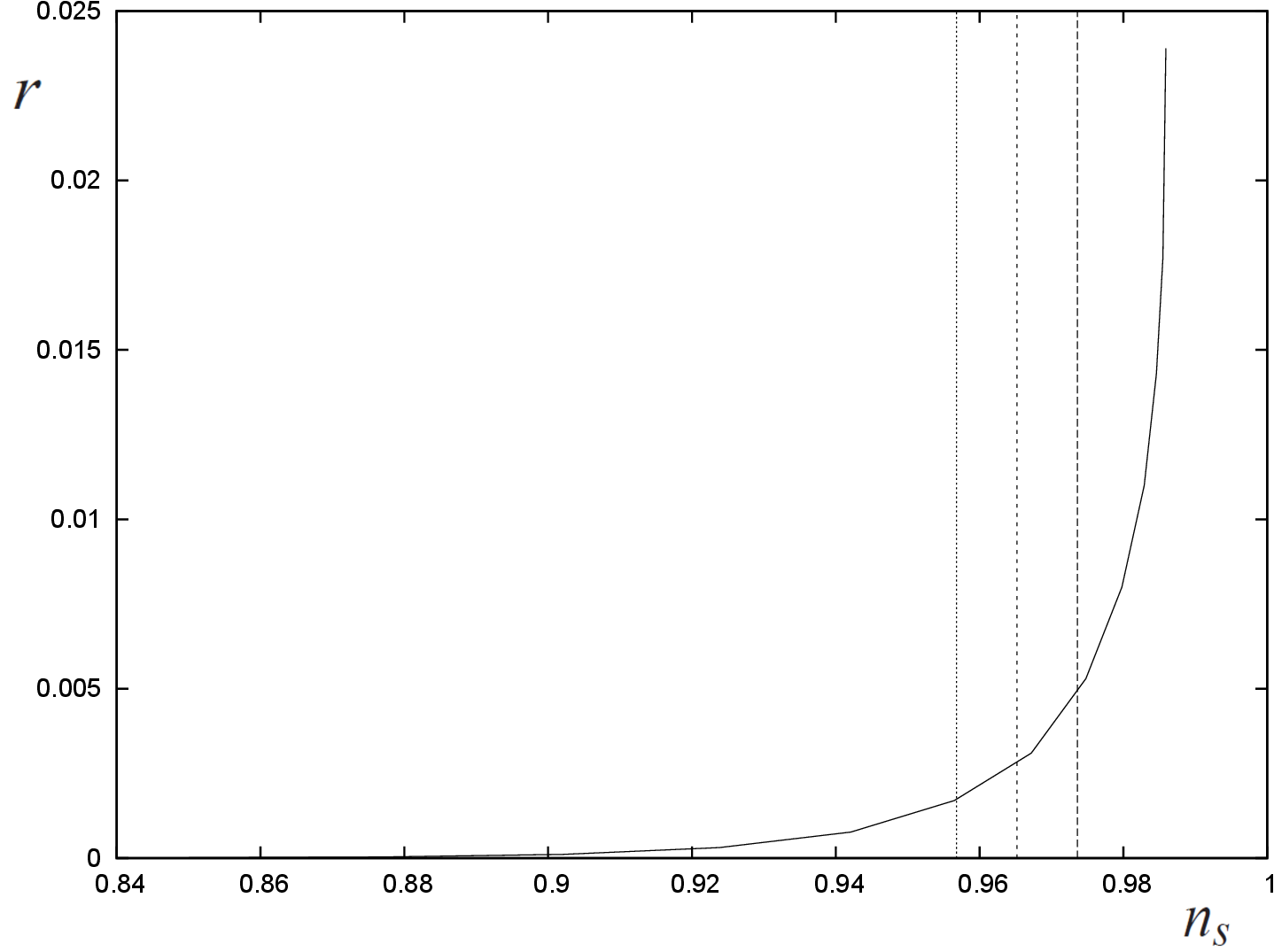}
\caption{$r$ as a function of $n_{s}$ for $M_{0} = 10^{10} \GeV$, showing how the model predicts a correlation of $n_{s}$ and $r$. The Planck best-fit value  and 2-$\sigma$ upper and lower bounds from $n_{s}$ are indicated.} 
\label{fig6}
\end{center}
\end{figure}

\begin{table}[htbp]
\begin{center}
\begin{tabular}{ |c|c|c|c|c|c|c| }
\hline
$\alpha$ 
& $n_{s}$ 
& $r$ & $\phi(57)$  & $\xi_{\phi}(m_{t})$ & $\xi_{\phi}(\mu_{c})$ & $\lambda_{\phi}(\mu_{c})$ 
\\
\hline
$1.7$ 
& $0.9857$ 
& $0.0239$ 
& $3.80$
& $2.43$
& $2.86$
& $0.0181$
\\
$1.5$ 
& $0.9852$ 
& $0.0177$
& $2.99$
& $2.66$ 
& $3.10$ 
& $0.0183$ 
\\
$1.4$ 
& $0.9843$ 
& $0.0143$
& $2.58$
& $2.86$
& $3.33$
& $0.0185$  
\\
$1.3$ 
& $0.9825$ 
& $0.0110$
& $2.17$
& $3.16$ 
& $3.68$ 
& $0.0187$ 
\\
$1.2$ 
& $0.9794$ 
& $0.0080$
& $1.78$
& $3.64$
& $4.24$ 
& $0.0189$ 
\\
$1.1$ 
& $0.9744$ 
& $0.0053$
& $1.42$
& $4.35$ 
& $5.06$
& $0.0193$ 
\\
$1.0$ 
& $0.9668$ 
& $0.0031$
& $1.09$
& $5.51$ 
& $6.41$ 
& $0.0198$ 
\\
$0.9$ 
& $0.9561$ 
& $0.0017$
& $0.82$
& $7.48$ 
& $8.69$ 
& $0.0204$ 
\\
$0.8$ 
& $0.9417$ 
& $0.00077$
& $0.59$
& $10.85$ 
& $12.60$ 
& $0.0211$ 
\\
$0.7$ 
& $0.9235$ 
& $0.00031$
& $0.42$
& $17.10$ 
& $19.81$
& $0.0221$ 
\\
$0.6$ 
& $0.9017$ 
& $0.00011$ 
& $0.29$
& $28.70$
& $33.18$
& $0.0231$ 
\\
$0.5$ 
& $0.8767$ 
& $0.000034$
& $0.20$
& $51.80$
& $58.80$
& $0.0244$ 
\\
$0.4$ 
& $0.8488$ 
& $0.0000085$
& $0.13$
& $104.99$ 
& $120.68$
& $0.0259$ 
\\
\hline
\end{tabular}
\caption{The scalar spectral index, $n_{s}$, the tensor-to-scalar ratio, $r$, the SM Higgs field $\phi$ at $N = 57$ (in units of $10^{9} \GeV$), the non-minimal coupling $\xi_{\phi}$ at $\mu = m_{t}$ (in units of $10^{3}$), $\xi_{\phi}$ at $\mu_{c}$, and the SM Higgs coupling $\lambda_{\phi}$ at $\mu_{c}$, as a function of $\alpha$ for $M_{0} = 10^{10} \GeV$. }
\end{center}
\end{table}

From the 2-$\sigma$ range of $n_{s}$, we find that for $M_{0} = 10^{10} \GeV$ the values of $\alpha$ are constrained to be in the range 
\be{r1} 0.91 < \alpha <  1.09  ~.\ee 
The corresponding range of $r$ is 
\be{r2} 0.0018 < r < 0.0050 ~,\ee 
with $r$ increasing as $n_{s}$ increases across the range allowed by Planck. Thus observation requires that $\alpha$ is close to 1 and so the renormalisation frame must be close to Prescription I. In particular, in the case of Prescription I with $\alpha = 1$, we find that the values of $n_{s}$ and $r$ are very close to the values predicted by the classical Higgs inflation, $n_{s} = 0.9667$ and $r = 0.0032$. In Appendix C we show that the reason $\alpha = 1$ produces results very close to the classical results for $n_{s}$ and $r$ is not because the quantum corrections to the potential are small, but because the  1-loop potential has the same functional form to leading order as the classical potential when $\alpha = 1$.

We also find that Prescription II ($\alpha = 0$) is completely excluded. We find no consistent solution that can produce inflation with the observed power spectrum when $\alpha \lae 0.1$.   

We find that the results for $n_{s}$ and $r$ are not very sensitive to $M_{0}$. In Table 2 we show the upper and lower bounds on $\alpha$ and $r$ for $M_{0}$ in the range $10^{4} \GeV$ to $5 \times 10^{11} \GeV$. The range starts to narrow as $M_{0}$ becomes larger than $10^{11} \GeV$. For $M_{0} \gae 10^{12} \GeV$ there is no solution for inflation on the metastable part of the potential. 

\begin{table}[htbp]
\begin{center}
\begin{tabular}{ |c|c|c|c|c| }
\hline
$M_{0} ({\rm GeV})$ 
& $\alpha_{lower}$ 
& $\alpha_{upper}$ & $r_{lower}$ & $r_{upper}$ 
\\
\hline
$10^{4}$ 
& $0.91$ 
& $1.08$ 
& $0.0017$
& $0.0047$
\\
$10^{6}$ 
& $0.90$ 
& $1.10$ 
& $0.0017$
& $0.0050$
\\
$10^{10}$ 
& $0.91$ 
& $1.09$ 
& $0.0018$
& $0.0051$
\\
$10^{11}$ 
& $0.94$ 
& $1.07$ 
& $0.0019$
& $0.0055$
\\
$5 \times 10^{11}$ 
& $0.97$ 
& $1.03$ 
& $0.0023$
& $0.0049$
\\
\hline
\end{tabular}
\caption{The upper and lower bounds on $\alpha$ and $r$ corresponding to the 2-$\sigma$ Planck bounds on $n_{s}$, as a function of $M_{0}$. }
\end{center}
\end{table}

We also note that inflation can be driven by a SM Higgs field value of the order of a TeV. For example, in Table 3 we show values of $n_{s}$ and $r$ for the case $M_{0} = 10^{4} \GeV$. In this case $\phi(N = 57)$ is in the TeV range.

We conclude that metastable Higgs Inflation can have a significantly larger or smaller value of $r$ than the classical Higgs Inflation prediction, and that these values are correlated with correspondingly larger or smaller values of  $n_{s}$. Renormalisation frames close to $\alpha = 1$, corresponding to Prescription I, are favoured by the observed range of $n_{s}$ from Planck.

\begin{table}[htbp]
\begin{center}
\begin{tabular}{ |c|c|c|c|c| }
\hline
$\alpha$ 
& $n_{s}$ 
& $r$ & $\phi(57)$ & $\xi_{\phi}(m_{t})$  
\\
\hline
$1.2$ 
& $0.9793$ 
& $0.0074$ 
& $3.36$
& $1.1 \times 10^{4}$
\\
$1.1$ 
& $0.9745$ 
& $0.0051$ 
& $3.03$
& $1.3 \times 10^{4}$
\\
$1.0$ 
& $0.9669$ 
& $0.0031$ 
& $2.71$
& $1.6 \times 10^{4}$
\\
$0.9$ 
& $0.9552$ 
& $0.0015$ 
& $2.40$
& $2.1 \times 10^{4}$
\\
$0.8$ 
& $0.9365$ 
& $0.0006$ 
& $2.13$
& $3.1 \times 10^{4}$
\\
\hline
\end{tabular}
\caption{Values of $n_{s}$, $r$, $\xi_{\phi}(m_{t})$ and $\phi$ at $N = 57$ (in TeV units) as a function of $\alpha$ for the case $M_{0} = 10^{4} \GeV$. In this case inflation is possible with a TeV scale Higgs expectation value.}
\end{center}
\end{table}

The LiteBIRD CMB polarisation satellite is predicted be able to measure $r$ to an accuracy $\delta r < 0.001$ \cite{litebird}. Therefore if a value of $r$ significantly larger or smaller than the classical prediction $r \approx 0.003$ is observed, and if $n_s$ is also found to be significantly larger or smaller than the Planck best-fit value, then metastable potential Higgs Inflation would be favoured. 

We also note that Early Dark Energy (EDE) models, which seek to resolve the $H_{0}$ tension, require larger values of $n_{s}$ than $\Lambda$CDM models, typically in the range 0.980-0.990 \cite{ede, ede2, ede3}. Metastable potential Higgs Inflation can be compatible with models with larger $n_{s}$, in which case $r$ could be significantly larger than the upper limit of 0.005 coming from the $\Lambda$CDM 2-$\sigma$ upper bound on $n_{s}$.

\subsection{Perturbative Unitary Violation} 

One issue that we have not yet discussed is perturbative unitarity violation due to the non-minimal coupling. This will be the same as in conventional Higgs inflation, with unitarity violated at tree-level in the inflationary background by Higgs scattering via graviton exchange at energies of the order of the background inflaton field $\overline{\phi} \sim M_{Pl}/\sqrt{\xi_{\phi}}$ \cite{fbu,lu}. If perturbative unitarity violation corresponds to true unitarity violation, then we would expect new physics with an energy scale $\overline{\phi}$ to modify the theory, and therefore we would expect non-renormalisable corrections to the potential scaled by inverse powers of $\overline{\phi}$. In this case Higgs Inflation would no longer be predictive.

However, perturbation theory breaks down at the energy of tree-level unitarity violation. Therefore it is possible and even likely that unitarity in Higgs scattering is conserved non-perturbatively.  Evidence that this is true was obtained by Tan and Willenbrock \cite{tw}, who showed that graviton scattering via the non-minimal coupling to gravity that is unitarity violating at tree level becomes unitarity conserving ("self-healing") once higher-order contributions are summed. This is further supported by the analyses in \cite{don} and \cite{calmet}. 

In this case there are no consequences for the Higgs potential or for the predictions of Higgs Inflation. The energy scale of quantum fluctuations during inflation, of order $ H \sim M_{Pl}/\xi_{\phi}$, is much smaller than the perturbative unitarity violation scale (i.e. the strong-coupling scale) $ \overline{\phi} \sim M_{Pl}/\sqrt{\xi_{\phi}}$, so there is no problem due to strongly-coupled Higgs scattering for the calculation of quantum fluctuations. There are also no non-renormalisable corrections to the potential due to new physics. The effect of the non-minimal coupling on the quantum corrections to the Higgs potential is fully taken into account by the propagator suppression factor, \eq{qc14}, which suppresses the physical Higgs contribution to the quantum corrections. Therefore the RG equations and the 1-loop CW correction to the potential used in our analysis are valid. Thus observational confirmation of the predictions of the model would also confirm non-perturbative unitarity conservation in Higgs Inflation.

\section{The significance of $\alpha$-frames with $\alpha$ close to 1}

$\alpha$-frames define just one set of possible renormalisation frames. One may therefore question the significance of these frames and whether there could be equally valid renormalisation frames that are not described by an $\alpha$-frame. Here we will argue that if the conformation transformation to the general renormalisation frame is a purely function of $\Omega$ and tends to 1 as $\Omega \rightarrow 1$, then $\alpha$-frames with $\alpha$ close to 1 (as preferred by the observed spectral index) will be a good approximation to any renormalisation frame that is a small deviation from Prescription I. That the conformal transformation is purely a function of $\Omega$ is a natural possibility, since $\Omega$ defines the non-minimal coupling in the Jordan frame. It is also likely that the conformal factor will tend to 1 as $\Omega$ tends to 1, since in this limit the non-minimal coupling in the Jordan frame becomes negligible.  

   To show that $\alpha$-frames with $\alpha$ close to 1 are a good approximation to any renormalisation frame close to Prescription I, we start by writing the transformation to a general renormalisation frame as 
\be{f1} \hat{\Omega}^{2} = \Omega^{2} F(\Omega)  ~\ee
where $F(\Omega)$ is a function of $\Omega$ which tends to 1 as $\Omega \rightarrow 1$. $F$ represents the deviation of the frame from Prescription I; for small deviations, $F$ will be close to 1. 
Without loss of generality, we can write $F$ in the form $F(\ln(\Omega))$. Since for renormalisation frames close to Prescription I we expect that $F \approx 1 $, and since $F \rightarrow 1$ as $\Omega \rightarrow 1$, we expect that $F(\ln(\Omega))$ can be expanded in the form 
\be{f2}  F(\ln(\Omega)) = 1 + k_{1}\ln(\Omega) + k_{2} \ln^{2}(\Omega) + ... \approx  1 + k_{1}\ln(\Omega)  ~,\ee 
where $k_{1}$ is small compared to 1.
Finally, defining $k_{1} = 2 (\alpha -1)$, where $\alpha$ is close to 1, we obtain 
\be{f4} F(\ln(\Omega)) \approx  1 + (\alpha -1)\ln(\Omega^{2})  \approx \exp \left( (\alpha - 1) \ln (\Omega^{2}) \right) = \Omega^{2(\alpha - 1)} ~.\ee
Therefore
\be{f5} \hat{\Omega}^{2} = \Omega^{2} F(\Omega) \approx \Omega^{2} \Omega^{2 (\alpha - 1)} \approx \Omega^{2 \alpha} ~.\ee 
Thus the $\alpha$-frames are likely to be good approximations to any renormalisation frame which is defined by a conformal transformation that is purely a function of $\Omega$ and which is a small deviation from the Prescription I frame.

\section{Conclusions} 

   Inflation via a metastable SM Higgs potential is an interesting and potentially significant possibility. It requires a change in the conventional assumptions of post-inflation cosmology, with a large increase in the Jordan frame effective Planck mass after the end of inflation without significant release of energy. The dynamics responsible for the Planck mass transition may come from an extension of the non-minimally coupled effective theory, as considered in \cite{pt1}, or may arise at a deeper level, as part of the UV completion of the theory. In any case, since the Planck mass transition is a necessity of inflation via a metastable Higgs potential, we can investigate the consequences of the model independently of the dynamics of the transition. We believe that the development of specific models for the Jordan frame Planck mass transition is well-motivated by the possibility of metastable SM Higgs Inflation. 

We find that quantum corrections can have a significant impact on the predictions of the model. Since the quantum corrections are purely SM corrections, they are well-defined, allowing the model to make clear predictions. The predictions depend upon the conformal frame in which the quantum corrections are calculated. We have generalised beyond the usual Prescription I and II renormalisation frames, since the correct renormalisation frame can only be determined by the UV completion of the theory. For the $\alpha$-frames we consider, we find that the observed range of $n_{s}$ restricts $\alpha$ to be close to the Prescription I frame, which corresponds to $\alpha = 1$. The model predicts a correlation between the values of $n_{s}$ and $r$, such that the 2-$\sigma$ Planck bounds on $n_{s}$ gives a possible range of $r$ between 0.0018 and 0.0050. Thus significantly larger or smaller values of $r$ than the classical value of 0.0032 are possible, which are correlated with larger or smaller $n_{s}$ and which can be tested by the detection of primordial gravitational waves by the next generation of CMB satellites.
In addition, the model can be compatible with the larger values of $n_{s}$ predicted by Early Dark Energy solutions to the Hubble tension, with correspondingly larger values of $r$.                                                                                                                                                                                                                                                                                                                                                                                                                                                                                                                                                                                                                                                                                                                                                                                                                                                                                                                                                                                                                                                                                                                                                                                                                                                                                                                                                                                                                                                                                                                                              It is interesting to consider that the observation of a large value for $r$ by LiteBIRD, combined with a large value for $n_{s}$, could be the signature of an alternative post-inflation cosmology.

   \renewcommand{\theequation}{A-\arabic{equation}}
 \setcounter{equation}{0}  

\section*{Appendix A: $N_{*}$ for Instant Reheating} 

In the Einstein frame, the energy density at the end of inflation is 
\be{rh1} \rho_{end} = V_{E} \approx \frac{\lambda_{\phi} M_{Pl}^{4}}{4 \xi_{\phi}^{2}}   ~.\ee 
Therefore, assuming instant reheating, the reheating temperature is 
\be{rh2} T_{R} = \left( \frac{30 \rho_{end}}{\pi^{2} g(T_{R})}\right)^{1/4} ~.\ee
The number of e-foldings at the pivot scale, $k_{*}$, is obtained from 
\be{rh3} \frac{2 \pi}{k_{*}} \left( \frac{a_{N}}{a_{0}} \right) \equiv  \frac{2 \pi}{k_{*}} \left(\frac{g(T_{0})}{g(T_{R})}\right)^{1/3} \left(\frac{T_{0}}{T_{R}}\right)e^{-N} = H^{-1} ~,\ee
where $T_{0}$ is the present CMB temperature, $g(T_{i})$ are the effective number of relativistic degrees of freedom, and $k_{*} = 0.05 \, {\rm Mpc}^{-1}$ is the Planck pivot scale. During inflation 
\be{rh4} H \approx \left(\frac{\rho_{end}}{3 M_{Pl}^{2}}\right)^{1/2}  ~.\ee 
Therefore we obtain 
 \be{rh5} N_{*} = \ln \left( \frac{2 \pi T_{0}}{k_{*}} \left(\frac{g(T_{0})}{g(T_{R})}\right)^{1/3}\left(\frac{\lambda_{\phi}}{\xi_{\phi}^{2}} \right)^{1/4} 
\left(\frac{\pi^{2} g(T_{R})}{1080} \right)^{1/4} \right)  ~\ee
Numerically we find 
\be{rh6} N_{*} = 62.6 + \frac{1}{4} \ln \left(\frac{\lambda_{\phi}}{\xi_{\phi}^{2}} \right) ~,\ee
where $\lambda_{\phi}$ and $\xi_{\phi}$ are calculated at $\mu = \mu_{c}$.  

For $M_{0} = 10^{10} \GeV$, using $T_{0} = 2.4 \times 10^{-13} \GeV$, $g(T_{0}) = 3.91$ and $g(T_{R}) = 106.75$,  we find that $N_{*}$ varies from 55.8 to 57.6 as $\alpha$ increases from 0.4 to 1.7. For $M_{0} = 10^{4} \GeV$ we find that $N_{*}$ varies from 56.9 to 57.4 as $\alpha$ varies from 0.8 to 1.2. Thus $N_{*} = 57$ is a good approximation to the number of e-foldings at the pivot scale.

\renewcommand{\theequation}{B-\arabic{equation}}
 \setcounter{equation}{0}  

\section*{Appendix B: One-loop Renormalisation Group Equations for the Non-Minimally Coupled Standard Model} 

The RG equations for the non-minimally coupled Standard Model are\footnote{There is a slight difference between the equations for $\xi_{\phi}$ between \cite{corr1} and \cite{corr2}, where \cite{corr1} has a factor $s^{2}(t)$ and \cite{corr2} has $s(t)$. This has only a very minor impact on the running. We follow \cite{corr2} here.} \cite{corr1,corr2} 
\be{rg1} \left(4 \pi \right)^{2} \frac{d g_{1}}{d t} = g_{1}^{3} \frac{\left(81 + s(t) \right) }{12}  ~.\ee 
\be{rg2} \left(4 \pi \right)^{2} \frac{d g_{2}}{d t} = -g_{2}^{3} \frac{\left(39 - s(t) \right) }{12}  ~.\ee 
\be{rg3} \left(4 \pi \right)^{2} \frac{d g_{3}}{d t} = -7 g_{3}^{3}   ~.\ee 
\be{rg4} \left(4 \pi \right)^{2} \frac{d y_{t}}{d t} = y_{t}\left( \left(\frac{23}{6} + \frac{2}{3} s(t) \right) y_{t}^{2} - 8 g_{3}^{2} - \frac{17}{12} g_{1}^{2} - \frac{9}{4} g_{2}^{2} \right) ~.\ee 
\be{rg5} \left(4 \pi \right)^{2} \frac{d \lambda_{\phi}}{d t} = \left( \left(6 + 18 s^{2}(t) \right) \lambda_{\phi}^{2} - 6 y_{t}^{4} + \frac{3}{8} \left(2 g_{2}^{4} + \left( g_{1}^{2} + g_{2}^{2} \right)^{2} \right) + 12 y_{t}^{2} \lambda_{\phi} - 3 g_{1}^{2} \lambda_{\phi} - 9 g_{2}^{2} \lambda_{\phi} \right)
~.\ee 
\be{rg6} \left(4 \pi \right)^{2} \frac{d \xi_{\phi}}{d t} = \left(\xi_{\phi} + \frac{1}{6} \right) \left( 6 \left(1 + s(t) \right) \lambda_{\phi} + 6 y_{t}^{2} - \frac{3}{2} g_{1}^{2} - \frac{9}{2} g_{2}^{2} \right)  ~,\ee
where $t = ln(\mu/\mu_{t})$ and $s(t)$ is given by \eq{qc14} with $\phi = m_{t}e^{t}$.

\renewcommand{\theequation}{C-\arabic{equation}}
 \setcounter{equation}{0}  

\section*{Appendix C: Why the $\alpha = 1$ predictions for $n_{s}$ and $r$ are almost the same as the classical predictions.}

At first sight it is surprising that Prescription I  ($\alpha = 1$) produces essentially identical predictions for $n_{s}$ and $r$ as the classical potential. Even though there is a suppression of $\phi$ dependence of the logarithms in the 1-loop CW potential, we would still expect to see some effect of the 1-loop potential. The reason that the deviation from the classical predictions for $n_{s}$ and $r$ are strongly suppressed at $\alpha = 1$ is not that the correction is very small, but that it has the same functional form as the tree level potential to leading order. As a result, 
even though there is a significant contribution to the effective potential from the 1-loop CW potential and the values of $\lambda_{\phi}$ and $\xi_{\phi}$ are significantly different from the classical values (for example, for $M_{0} = 10^{10} \GeV$ we find $\lambda_{\phi}(\mu_{c}) = 0.02$ and $\xi_{\phi}(m_{c}) = 6.4 \times 10^{3}$, compared to the classical values $\lambda_{\phi} = 0.126$ and $\xi_{\phi} = 1.8 \times 10^{4}$), the predictions for $n_{s}$ and $r$ are hardly modified when expressed as a function of $N$. 

To illustrate this, we will consider only the t-quark contribution to the CW potential, which is the largest contribution during inflation. After renormalising in the $\alpha$-frame and then transforming to the Einstein frame (with effective Planck mass $M_{0}$), the potential is given by 
\be{cf1} \overline{V}_{E}(\phi) = \frac{\lambda_{\phi} \phi^{4}}{4 \left(1 + \frac{\xi_{\phi} \phi^{2}}{M_{0}^{2}} \right)^{2}} - \frac{3 y_{t}^{4} \phi^{4}}{4 \left(1 + \frac{\xi_{\phi} \phi^{2}}{M_{0}^{2}} \right)^{2}}
\left[ \ln \left( \frac{y_{t}^{2} \phi^{2}/2}{\mu_{c}^{2} \left(1 + \frac{\xi_{\phi} \phi^{2}}{M_{0}^{2}} \right)^{\alpha} } \right) - \frac{3}{2} \right]  ~.\ee
During inflation, we have $\xi_{\phi} \phi^{2} \gg M_{0}^{2}$. Expanding $\overline{V}_{E}(\phi)$ in $M_{0}^{2}/(\xi_{\phi} \phi^{2})$, we obtain 
\be{cf2}  \overline{V}_{E}(\phi) = A \left(1 - \frac{B K}{A} \right)\left(
1 - \frac{2 \left(1 - B/A\right)}{\left(1 - BK/A\right)} \left(\frac{M_{0}^{2}}{\xi_{\phi} \phi^{2}}\right) + \frac{B (1 - \alpha)}{A(1 - BK/A)} \ln \left(\frac{M_{0}^{2}}{\xi_{\phi} \phi^{2}}\right)  + O \left(\frac{M_{0}^{2}}{\xi_{\phi} \phi^{2}}\right)^{2}\right)      ~,\ee
where $A = \lambda_{\phi}^{2} M_{0}^{4}/4 \xi_{\phi}^{2}$, $B  = 3 y_{t}^{4} M_{0}^{4}/4 \xi_{\phi}^{2}$ and $K = \ln[y_{t}^{2} M_{0}^{2}/2 \xi_{\phi} \mu_{c}^{2}] -3/2$. Therefore to leading order in $M_{0}^{2}/\xi_{\phi}\phi^{2}$ the potential has the form 
\be{cf3} \overline{V}_{E} = V_{0}\left( 1 - c \frac{M_{0}^{2}}{\xi_{\phi} \phi^{2}} +  d (1 - \alpha) \ln \left[ \frac{M_{0}^{2}}{\xi_{\phi} \phi^{2}} \right]   \right) \ee 
where $V_{0}$, $c$  and $d$ are constants. The classical potential to leading order has the form 
\be{cf3} \overline{V}_{E,\,cl} = V_{0,\,cl}\left( 1 -  \frac{2 M
_{0}^{2}}{\xi_{\phi} \phi^{2}} \right) ~.\ee 
If $\alpha \neq 1$ then the leading-order effective potential has a different functional form from the classical potential. But if $\alpha = 1$  then the form of the potential is the same, only the values of the constants $V_{0}$ and $c$ are different. However, the predictions for $n_{s}$  and $r$ are independent of the constants $V_{0}$ and $c$ when they are expressed as a function of $N$. Therefore if $\alpha = 1$ then the predictions of the classical and quantum potentials are the same to leading order. This is confirmed numerically, where there is very little difference between the predictions for $n_{s}$ and $r$. This is not because the quantum corrections to the potential in Prescription I are small, but because they do not modify the functional form of the potential to leading order in $M_{0}^{2}/\xi_{\phi} \phi^{2}$.

\end{document}